\documentclass{elsartx}
\usepackage[square,comma]{natbib}

\def\al{\alpha}
\def\be{\beta}
\def\ga{\gamma}
\def\de{\delta}

\def\ka{\kappa}
\def\la{\lambda}

\def\rh{\rho}

\def\ph{\phi}

\def\ps{\psi}

\def\Ga{\Gamma}
\def\De{\Delta}

\def\La{\Lambda}

\def\cl{{\mathcal L}}

\def\fr#1#2{{{#1} \over {#2}}}

\def\frac#1#2{{\textstyle{{#1}\over {#2}}}}

\def\lsim{\mathrel{\rlap{\lower4pt\hbox{\hskip1pt$\sim$}}
    \raise1pt\hbox{$<$}}}
\def\gsim{\mathrel{\rlap{\lower4pt\hbox{\hskip1pt$\sim$}}
    \raise1pt\hbox{$>$}}}
\def\sqr#1#2{{\vcenter{\vbox{\hrule height.#2pt
         \hbox{\vrule width.#2pt height#1pt \kern#1pt
         \vrule width.#2pt}
         \hrule height.#2pt}}}}

\def\prt{\partial}

\def\ol#1{\overline{#1}}

\newcommand{\beq}{\begin{equation}}
\newcommand{\eeq}{\end{equation}}
\newcommand{\bea}{\begin{eqnarray}}
\newcommand{\eea}{\end{eqnarray}}
\newcommand{\bit}{\begin{itemize}}
\newcommand{\eit}{\end{itemize}}
\newcommand{\rf}[1]{(\ref{#1})}

\begin{document}

\begin{frontmatter}

\title{Spontaneous Lorentz Violation and Nonpolynomial Interactions}

\author{B.\ Altschul and V.\ Alan Kosteleck\'y}

\address{Physics Department, Indiana University,
Bloomington, IN 47405}

\date{IUHET 481}
\journal{Physics Letters B}

\begin{abstract}
Gauge-noninvariant vector field theories 
with superficially nonrenormalizable nonpolynomial interactions
are studied. 
We show that nontrivial relevant and stable theories
have spontaneous Lorentz violation, 
and we present a large class of asymptotically free theories.
The Nambu-Goldstone modes of these theories
can be identified with the photon,
with potential experimental implications.
\end{abstract}

\end{frontmatter}

\section{Introduction}

Experiments show that nature is well described 
at presently accessible energies 
by two field theories:
the Standard Model (SM) of particle physics,
and Einstein's General Relativity.
These are expected to arise as the low-energy limit
of a fundamental theory of quantum gravity 
at the Planck scale,
$M_P \simeq 10^{19}$ GeV.
The discrepancy between $M_P$ and attainable energies
makes experimental signals from this underlying theory 
difficult to identify,
but one promising class of observables 
involves violations of Lorentz symmetry
arising from new physics at the Planck scale.\footnote{%
For a variety of recent reviews 
see, e.g., Refs.\ \cite{cpt04,rb,dm,almm,ss}.}

An interesting and challenging issue 
that has received little attention to date 
is the extent to which Lorentz violation 
might be generic or even ubiquitous 
in prospective fundamental theories.
The present work initiates a study of this issue.
For definiteness,
we focus attention on
the elegant possibility that Lorentz symmetry 
is spontaneously broken in the underlying theory
\cite{ks}.
The basic idea is that interactions in the underlying theory 
induce nonzero vacuum expectation values for 
one or more Lorentz tensors,
which can be regarded as background quantities
in the vacuum throughout spacetime.

The analysis in this work adopts the methods 
of lagrangian-based quantum field theory,
in which observable effects of Lorentz violation
are described by an effective low-energy field theory 
\cite{kp,ck,akgrav},
and it assumes that the issue of obtaining the required hierarchy
\cite{kp,cpsuv,almm}
for the associated coefficients for Lorentz violation
can be addressed.
In this language,
the fundamental theory may have many types of fields and interactions,
including ones that are nonrenormalizable
at the level of the effective low-energy theory.
A comprehensive study of the likelihood of Lorentz violation
at this level appears infeasible at present.
However,
in Lorentz-invariant scalar field theories, 
certain nonpolynomial 
and hence superficially nonrenormalizable interactions 
have been shown to be relevant
in the sense of the renormalization group (RG) 
by studying the natural cutoff dependences 
of the coupling constants 
\cite{hh}. 
The gaussian fixed point 
of the RG flow is ultraviolet stable
along certain directions in the parameter space of interactions,
and these directions correspond 
to nontrivial asymptotically free theories. 
Here,
we exploit this idea 
by generalizing the scalar analysis 
to the case of vector fields
and investigating the occurrence 
of spontaneous Lorentz violation in the resulting theories.

The prototypical field theories for Lorentz violation 
with a vector field $B^\mu$ are the so-called bumblebee models 
\cite{ks2}.\footnote{%
Recent literature includes
Refs.\ \cite{akgrav,kleh,jme,kt,jm,bg,cfmn,bkssb,cl,gjw,hbh,bp,ems}.}
These involve a gauge-noninvariant potential $V(B^\mu B_\mu)$ 
that has a minimum at nonzero $B^\mu$ 
inducing spontaneous Lorentz violation.
We consider a generic model of this type
with a conventional Maxwell-type kinetic term
and an arbitrary nonpolynomial potential,
as might arise from a fundamental theory,
and we take RG relevance of the interactions 
and stability of the associated quantum field theory 
as practical criteria determining acceptable models
for our study.
These assumptions make it possible to address
the ubiquity of Lorentz violation
in a definite context.
Surprisingly,
we find that consistent stable relevant theories 
of this type 
{\it must} have spontaneous Lorentz violation
and that a large class of such theories arises from
superficially nonrenormalizable bumblebee models.
Moreover,
these theories naturally contain 
Nambu-Goldstone (NG) modes 
\cite{ng}
associated with spontaneous Lorentz violation
that can be identified with the photon,
a result with potential experimental consequences.

To perform the analysis,
we study the flow in the Wilson formulation of the RG
\cite{wk}, 
in which the theory is considered with a momentum cutoff.
An analysis with a more general cutoff is also possible
\cite{vp},
through the use of the Polchinski formulation of the RG
\cite{jp}.
Since any Lorentz violation in nature is a weak effect, 
we restrict attention 
to the linearized form of the RG transformation, 
in which only terms that are first-order 
in the interaction are retained. 
The literature for the Lorentz-invariant scalar case
contains some discussion 
about the persistence of the nonpolynomial interactions 
when the full nonlinear RG is considered
\cite{tm}.
However,
a nonperturbative demonstration to the contrary would require
showing that the nonpolynomial potentials 
can be expanded as a sum of an infinite number 
of irrelevant RG modes,
a challenging task. 
Moreover,
there is evidence for persistence:
in the limit where the number of scalar-field components is large,
it is known that the RG equations for the nonpolynomial modes
can be integrated into a region far from the fixed point
\cite{hg}.
This suggests that the novel potentials 
exist outside the linearized regime 
for finite-component fields as well.
Other generalizations of the original results include
Refs.\ \cite{ab,vb}
and a study of the impact of Lorentz violation 
on asymptotically free scalar and spinor field theories
\cite{ba}.
No evidence exists for relevant nonpolynomial theories
involving spinor fields, 
but a leading-order analysis 
shows that Lorentz violation is a prerequisite 
for their existence, 
a conclusion compatible with the results for vector fields
obtained below. 

\section{Running the bumblebee}

Consider a theory for a vector-valued 
`bumblebee' potential field $B^{\mu}$
with Lagrange density 
\beq
\cl=-\frac 14 B^{\mu\nu}B_{\mu\nu}+ V(B^\mu B_\mu),
\label{lag}
\eeq
where $B_{\mu\nu}= \prt_\mu B_\nu - \prt_\nu B_\mu$ 
is the field strength. 
The potential $V$ 
is assumed to be representable as a power series in $B^2$,
and it violates gauge invariance.
For the simple case $V= m^2 B^2/2$, 
the theory describes a free massive vector boson.
Potentials such as $V= \la (B^2 - b^2)^2/4$ for constant $b^2$
produce bumblebee models describing spontaneous Lorentz violation.
Here,
we consider a theory with more general nonpolynomial $V$.
In what follows,
we introduce a momentum cutoff $\La$
representing the only scale in the system, 
and we insert appropriate powers of $\La$ in $V$ 
to render dimensionless all the couplings
\cite{hh}.

To study the RG flow for the theory \rf{lag},
it is convenient to Wick rotate the variables 
and operate in euclidean space.
For scalar fields, 
the euclidean RG equations are known 
\cite{wh}.
The RG calculations for the Wick-rotated vector field 
parallel those for an SO(4) multiplet of four scalar fields, 
except for a minor modification
arising from the structure of the kinetic term. 
At tree level,
the transversality of the bumblebee kinetic term
ensures there are only three propagating modes
contained in the four-component field $B^\mu$.
This feature remains true in the RG analysis
because no kinetic contributions to the two-point function 
for the fourth mode can arise.
The relevant diagrams either have both external legs
on the same vertex,
yielding a tadpole and no kinetic contribution,
or they have an internal line for the fourth mode,
which vanishes.
Since only three modes propagate rather than four,
we must replace the zero-separation scalar propagator
\beq
\De^{jk}_F(0)=\int_{|p|<\La}\fr{d^4p}{(2\pi)^4}
\fr {\de^{jk}}{p^2}
\eeq
with the transverse propagator
\bea
D^{\mu\nu}_F(0) & = & 
-\int_{|p|<\La}\fr{d^4p}{(2\pi)^4}
\fr {(\de^{\mu\nu}-p^\mu p^\nu/p^2)}{p^2} 
= -\frac 34 \De^{\mu\nu}_F(0) 
= - \fr {3\La^2}{64\pi^2} \de^{\mu\nu}.
\label{last}
\eea
However, 
the extra factor of $3/4$ relative to the scalar case 
contributes only to the overall normalization of the vector field.

The asymptotically free solutions 
of the linearized RG equations are
\cite{hh}
\beq
V_\ka (B^{2})=
g\La^{4} 
\left[ M (\ka -2; 2; z)-1
\right].
\label{sols}
\eeq
Here, 
$M(\al;\be;z)$ is the confluent hypergeometric (Kummer) function
\cite{as},
\beq
M(\al;\be;z)=
1+\fr{\al}{\be}\fr{z}{1!}+\fr{\al(\al+1)}
{\be(\be+1)}\fr{z^{2}}{2!}+\cdots,
\label{eq-Mdef}
\eeq
with $z=-32\pi^2 B^2/3\La^2$.
The parameter $\ka$ in Eq.\ \rf{sols}
describes the growth of the coupling constant $g$ 
when the cutoff scale $\La$ is changed. 
If all modes with momenta in the range 
$\La_1 < |p| < \La_0$ 
are integrated out of the theory 
and the fields rescaled accordingly, 
then the renormalized $g$ shifts to 
$g(\La_0/\La_1)^{2\ka}$.
Asymptotically free theories must have $\ka>0$, 
and only these theories have nontrivial continuum limits.
It is convenient to parametrize the coupling $g$ as
\beq
g=\fr c {\ka -2},
\eeq
so the sign of $c$ gives the sign of the slope 
of $V_\ka$ at $z=0$.

Substantial additional complexities arise 
when the theory in euclidean space is reconverted 
to Minkowski spacetime. 
For the scalar-field case, 
$\ph^2=\sum_{j=1}^4 (\ph^j)^2$ 
is guaranteed to be positive, 
but the analogous quantity $-B^2= -B^\mu B_\mu$
for vector fields
can be either positive (spacelike $B^\mu$) 
or negative (timelike $B^\mu$).
This complicates the analysis of the stability 
of these theories.
Furthermore, 
any nontrivial, stable, asymptotically free theory 
of this type necessarily involves spontaneous Lorentz breaking. 
This follows because $z$ can now be positive or negative.
If $V_\ka$ either increases or decreases at $z=0$, 
then there is a state with nonzero $B^\mu$ 
having lower energy than a state with $B^\mu=0$, 
so $B^\mu$ develops a Lorentz-violating vacuum expectation value. 
If instead $V_\ka$ has a vanishing derivative at $z=0$,
then $c=0$, the potential vanishes identically, 
and the theory is trivial.

\section{Stability analysis}

To determine which $\ka$ correspond to stable theories, 
we examine the asymptotic behavior of $V_{\ka}$ 
as $z\rightarrow+\infty$ and $z\rightarrow-\infty$. 
For large positive $z$, 
the asymptotic formula
\beq
M(\al;\be;z)\approx \fr{\Ga(\be)z^{\al-\be}e^z} {\Ga(\al)}
\label{masym}
\eeq
holds,
with all corrections being suppressed by powers of $z^{-1}$. 
Also useful is the Kummer formula
\beq
M(\al;\be;-z)=e^{-z}M(\be-\al;\be;z),
\eeq
which is an exact relation.

Consider first the case of a spacelike expectation value 
for $B^\mu$, 
so that the minimum of $V_\ka$
has $-B^2 >0$ and hence $z>0$.
This parallels the case 
of positive vacuum expectation value for $\ph^2$ 
\cite{hh}.
Suppose $c<0$. 
The potential is then decreasing with $z$ at $z=0$, 
so if $V_\ka$ diverges to positive infinity for large $z$
then at least one stable minimum must exist for $z>0$.
Equation \rf{masym} shows $V_{\ka}$ indeed diverges 
as $z\rightarrow+\infty$, 
and its sign is determined by the sign of 
$c/(\ka -2) \Ga(\ka -2)$
or, equivalently, 
by the sign of $-\Ga(\ka -1)$. 
This is positive when $\ka$ lies 
in one of the open intervals 
$(0,1)$, $(-2,-1)$, $(-4,-3)$, etc. 
Since a nontrivial theory must have $\ka>0$, 
the only relevant range is $0<\ka<1$.
In contrast,
if $c>0$, 
then there are no stable potentials 
with local minima on the positive $z$-axis because
either $V_\ka \rightarrow-\infty$ 
as $z\rightarrow+\infty$
or $V_\ka$ increases monotonically.

The potentials $V_\ka$ of interest 
that generate a spacelike expectation value for $B^\mu$ 
are therefore those with $c<0$ and $0<\ka<1$,
corresponding to $g>0$ in the range $|c/2|<g<|c|$.
The stability of $V_\ka$ for $z<0$
must however also be verified.
For negative $z$,
\beq
V_\ka (B^2)= g
\left[
e^{-|z|} M (4- \ka; 2; |z| ) -1
\right].
\label{Vneg}
\eeq
In the range $0<\ka<1$, 
the hypergeometric function grows faster with $|z|$ than
$e^{|z|}=M(2;2;z)$.
Since $g>0$,
it follows that $V_\ka$ is positive for all negative $z$. 
These theories are therefore stable,
with a spacelike expectation value for $B^\mu$
and spontaneous Lorentz violation. 

Next,
consider the case of a timelike expectation value 
for $B^\mu$, 
for which the vacuum has $-B^2 <0$ and $z<0$.
Any stable theories of this type must have $\ka\geq 1$,
and also $c>0$ is required for stability as $z\rightarrow+\infty$.
Moreover,
as $z\rightarrow-\infty$
the asymptotic behavior of the hypergeometric function 
is determined by
\beq
\fr c {\ka -2}e^{-|z|} M \left(4- \ka; 2; |z| \right)
\approx
\fr c {\ka -2}
\fr {\Ga(2)|z|^{2- \ka}} {\Ga (4- \ka)}.
\eeq
For $1\leq\ka <2$, 
this diverges to negative infinity for large negative $z$,
so the potentials in this range are unstable.
For $\ka=2$,
the potential vanishes. 
For $\ka>2$, 
we find $V_\ka \rightarrow -c/(\ka -2)$
as $z\rightarrow -\infty$, 
since $|z|^{2-\ka}\rightarrow 0$. 
The theory is therefore stable 
if there exists a $z<0$ for which 
$V_\ka\leq-c/(\ka -2)$,
which occurs if and only if
$M (4- \ka; 2; |z|)$ has a root.
This function cannot have a root unless $\ka>4$
because otherwise all the terms in the sum \rf{eq-Mdef} are positive. 
However,
$M(\al;2;|z|)$ does indeed have a root 
for $\al$ sufficiently large and negative.
The absolute value of the smallest root decreases
as $\al$ becomes more negative
\cite{jel}. 
In fact,
there is a root for any $\al<0$,
i.e., any $\ka>4$:
if $-1<\al<0$, 
then the asymptotic value of $M(\al;2;|z|)$ is negative
and so it must possess a root.

In summary,
we find that theories having potentials $V_\ka$ with $0< \ka<1$ 
are stable,
with minima lying at spacelike values of $B^{\mu}$. 
Theories with $1\leq\ka<2$ and $2<\ka\leq 4$ are unstable,
while the case $\ka=2$ is trivial. 
Stability is restored for $\ka>4$,
and the vacuum value of $B^{\mu}$ becomes timelike. 
A timelike vacuum value for $B^{\mu}$ may in fact be favored 
because the potentials leading to this form of symmetry breaking 
are more relevant.

The potentials for $0<\ka<1$ are discussed in
Ref.\ \cite{hh}. 
The hypergeometric functions 
$M(\ka-2;2;z)$ 
have minima with $z<10$ for nearly all $\ka$,
and the values of the Kummer functions 
at these minima are typically also less than 10.
However, 
as $\ka\rightarrow 1$
the potential evolves into an unstable inverted parabola, 
and so both the location of the minimum 
and its value diverge in this limit.
In the timelike range $\ka>4$, 
the potential may possess multiple local minima 
at negative values of $z$. 
However, 
the exponential damping factor $e^{-|z|}$ 
in Eq.\ \rf{Vneg} ensures that  
the one with the smallest $|z|$ is always the global minimum.
As $\ka$ increases,
the wavelength of the oscillations in $V_\ka (z)$ decreases,
and the location of the minimum 
is pushed to smaller values of $|z|$. 
This location may be calculated numerically, 
and it is roughly given by 
$z_{\rm min}\approx -6/(\ka -3)$. 
The value $V_{\ka, {\rm min}}$ of $V_\ka$ at $z_{\rm min}$ is
consistently close to 
$V_{\ka, {\rm min}} \approx 
-0.1g \La^2 (\ka -3)$. 

\section{Features and implications}

We have shown that $B^\mu$ 
must develop a Lorentz-violating vacuum expectation value 
in any nontrivial stable theory. 
There are many potential implications
of this scenario.
An immediate one concerns the interpretation 
of the excitations about the vacuum.
Denoting the vacuum value as 
$\langle B^\mu\rangle=b^\mu$,
we may parametrize $B^\mu$ as
\beq
B^\mu=(1+\rh)b^\mu+A^\mu,
\label{bdef}
\eeq
where $b^\mu A_\mu =0$. 
Defining
$F_{\mu\nu}=\prt_\mu A_\nu -\prt_\nu A_\mu$,
the kinetic term for $A^\mu$ is found to be 
$-\frac 1 4 F^{\mu\nu}F_{\mu\nu}$. 
Moreover,
at lowest order 
only fluctuations in $\rh$ 
cause changes in the potential term in the energy,
so there is no mass term for $A^\mu$.
We therefore can identify $A^\mu$ with the photon field.

The notation in Eq.\ \rf{bdef}
is chosen to match that of 
Ref.\ \cite{bkssb},
which provides a general description 
of the NG modes
associated with spontaneous Lorentz violation
and presents the complete effective action for $A^\mu$
in various spacetimes.
In this context,
excitations around the vacuum of the field $A^\mu$ 
are the NG modes associated 
with spontaneous Lorentz breaking, 
while vacuum excitations of $\rh$
are the NG modes for spontaneous diffeomorphism breaking. 
The masslessness of the photon follows directly
from this interpretation
as a consequence of the breaking 
of Lorentz invariance\footnote{%
In the context of electrodynamics 
without physical Lorentz violation,
this interpretation has a long history.
See, for example,
Ref.\ \cite{dhfbn}.}
rather than the existence of gauge symmetry.
The effective action also contains higher-order corrections 
to conventional electrodynamics
that could be sought in experimental tests. 
The superficially nonrenormalizable couplings are suppressed by
powers of $\La$ and vanish in the continuum limit.

Since at leading order the potential $A^\mu$ 
satisfies the orthogonality condition $b^\mu A_\mu=0$,
the equivalent conventional electrodynamics
must be defined in an axial or generalized axial gauge. 
One check on the quantum equivalence 
between electrodynamics and the theory \rf{lag}
at leading order is provided by a comparison 
of the corresponding transverse propagators. 
In fact, 
the euclidean propagator for electrodynamics
subject to the gauge condition $b^\mu A_\mu=0$ is\footnote{%
For a discussion of the propagator in axial gauges see,
for example, Ref.\ \cite{gl}.}
\bea
D^{\mu\nu}_F(x-y)&=& 
-\int\fr{d^4 p}{(2\pi)^{4}}
e^{-ip\cdot (x-y)} 
\fr{1}{p^{2}}\Bigg[\de^{\mu\nu}
+\fr{b^2}{(b\cdot k)^2}p^\mu p^\nu
-\fr{b^\mu p^\nu +b^\nu p^\mu }{(b\cdot k)}\Bigg].
\nonumber\\
\eea
Within the subspace of propagating modes, 
this is equivalent to the corresponding propagator 
for the theory \rf{lag}. 

The formulation of the theory \rf{lag} 
involves three propagating modes,
and three modes also appear after the spontaneous Lorentz breaking.
Two are photon modes contained in $A^\mu$. 
The third mode is massive, 
with excitations that change the value of $B^2$
and hence the potential.
The curvature at the global minimum 
of the potential determines the mass 
of the fluctuations of $B^2$ about its vacuum value 
$b^\mu b_\mu = b^2$. 
This curvature is always proportional to $g\La^{2}$. 
However,
$g$ must be small for the linearized calculations to be valid, 
while the natural physical cutoff scale $\La$ 
is expected to be very large, 
possibly of the order of $M_P$. 
It is therefore reasonable to expect  
that the particles associated with the fluctuations of $B^2$ 
are unobservable in a low-energy theory.
A large value of $\ka$ implies both small
$b^2$ and a large mass for these particles, 
so weak Lorentz violation is naturally associated 
with unobservability of the massive mode.

For $\ka>4$,
the potential remains finite 
at infinite positive timelike values of $B^{2}$.
The energy density required to shift the expectation value of
the field arbitrarily far from its vacuum value 
therefore remains finite.
However, 
if $\La$ is of the order of $M_P$,
this energy density is proportional to $gM_P^4$.
This is suppressed by only one power of $g$ relative 
to the naive scale of the cosmological constant. 
The energy density required to generate 
these large field values exists only in the early Universe, 
when high-temperature corrections are expected to restore 
the broken Lorentz symmetry. 

Another interesting feature of the timelike case
is the inverse variation with $\ka$ of the 
locations of the minima of the $V_\ka$ potentials,
$z_{\rm min}\approx -6/(\ka -3)$. 
Suppose the underlying theory
contains a sum of nonpolynomial interactions $V_\ka$ 
with values of $\ka$ ranging to a maximum $\ka_{\rm max}$
and with coefficients for the different $V_\ka$ potentials 
controlled by the details of the fundamental physics. 
When the spontaneous Lorentz breaking 
is studied in a lower-energy effective field theory
with a smaller value of the cutoff, 
then the effective potential is dominated 
by those $V_\ka$ with $\ka$ in the vicinity of $\ka_{\rm max}$ 
because these potentials grow 
the most rapidly as the cutoff decreases. 
The magnitude of the Lorentz-violating vector $b^\mu$ 
is then proportional to $1/\sqrt{\ka_{\rm max}}$. 
Since $\ka_{\rm max}$ represents the maximum 
in a potentially large collection of $\ka$ values, 
it can naturally be big. 
This could provide a partial explanation 
for the small size of any Lorentz violation in nature.

Additional physical implications of our scenario 
can be explored by extending the theory \rf{lag} 
to include couplings between $B^\mu$ and one or more other fields. 
For example,
introducing a Dirac fermion field $\ps$ 
offers various possibilities for interactions
between $B^\mu$ and fermion bilinears.
Note that gauge-noninvariant couplings are acceptable here,
unlike the usual case of quantum electrodynamics (QED),
because the initial theory \rf{lag} has no gauge invariance. 
Whatever the nature of the other fields being introduced,
Lorentz-violating terms for them appear 
following spontaneous Lorentz violation
when $B^\mu$ is replaced with its vacuum value $b^\mu$
in the interactions.
If the additional fields are identified with ones 
in the SM or in gravity,
the resulting Lorentz violation 
is contained in the Standard-Model Extension
(SME)
\cite{ck,akgrav}.
For example,
a simple choice of interaction is 
$\cl_a \propto B_\mu \ol\ps\ga^\mu\ps$,
paralleling the usual QED current coupling.
When $B_\mu$ acquires a vacuum value,
this interaction generates 
the usual coupling of $A_\mu$ to the current
along with a coefficient for Lorentz violation 
of the $a_\mu$ type in the minimal Lorentz-violating QED extension.
For a single fermion, 
a constant coefficient $a_\mu$ is unobservable,
but when fermion flavor changes are present
coefficients of this type
can produce observable effects 
\cite{ak,kmnu}.
More exotic couplings could also be countenanced,
such as an axial-vector coupling 
$\cl_b \propto B_\mu\ol\ps\ga_5\ga^\mu\ps$
\cite{kleh}. 
When Lorentz symmetry is spontaneously broken, 
this induces a coefficient for Lorentz violation 
of the $b_\mu$ type in the SME,
along with a remnant interaction involving the fluctuations
about $b^\mu$.
Note that all these couplings are known to be 
renormalizable at one loop
\cite{kpl}.
If the massless excitations are identified 
with the photon $A^\mu$ as above,
then a novel coupling to the photon arises.

Many other types of couplings for $B^\mu$
can also be considered.
Although beyond our present scope,
it would be of definite interest to explore features
introduced by derivative couplings 
within this framework.
These could provide insight about the structure
of higher-derivative terms in the effective action
and hence about the causality and stability 
of theories with Lorentz violation
\cite{kleh},
and they could have a bearing 
on predicted Lorentz-violating effects
within the photon sector
\cite{km}.
It would also be of interest to investigate
gravitational couplings, 
including background spacetimes
\cite{akgrav}.
Incorporating gravity typically makes 
RG calculations of the type adopted here impractical, 
but comparatively simple cases 
such as conformally flat backgrounds
may be tractable. 
In any event,
the occurrence of spontaneous Lorentz violation
as a necessary feature in the above stable 
and relevant theories with nonpolynomial potentials
suggests that Lorentz violation 
might indeed be generic in a large class of underlying 
theories at the Planck scale.

This work was supported in part
by DOE grant DE-FG02-91ER40661
and NASA grant number NAG3-2914.

\end{document}